\documentclass[12pt]{article}
\usepackage{geometry}                
\usepackage{graphicx}
\usepackage{amssymb}
\usepackage{cite}
\usepackage{amsmath}
\usepackage{bm}

\begin{document}

\title{~\\
\bf 
Higgs Decay into Two Photons through the  {\boldmath $W$}-boson Loop:
No Decoupling\\ in the  {\boldmath $m_W\to 0$} Limit
 }
\author{M.\ Shifman,$^a$ A.\ Vainshtein,$^a$ M.B.\ Voloshin,$^{\,a,b}$ and V.\ Zakharov$^{\,b,c}$\\[2mm]
{\small $^a$ ~\sl William I. Fine Theoretical Physics Institute, University of Minnesota,}\\[-1mm]
{\small \sl 
Minneapolis, MN 55455, USA}\\[-1mm]
{\small $^b$ ~\sl Institute for Theoretical and Experimental Physics, B. Cheremushkinskaya 25, }\\[-1mm]
{\small \sl 
Moscow, 117218, Russia}\\[-1mm]
{\small $^a$ ~\sl Max-Planck-Institut f\"ur Physik, F\"ohringer Ring 6, }\\[-1mm]
{\small \sl 
80805 M\"unchen, Germany}
}

\date{}                                           

\maketitle
\thispagestyle{empty}

\vspace{-10.9cm}

\begin{flushright}
FTPI-MINN-11/21\\ 
UMN-TH-3011/11\\
\end{flushright}

\vspace{8.5cm}

\begin{abstract}
We reanalyze  the $W$-boson loop in the amplitude of the Higgs decay into two photons
to show the absence of decoupling in the limit of massless $W$ bosons,
$m_W \to 0$. The Higgs coupling to longitudinal polarizations survive in
this limit and generates a nonvanishing contribution in the  $H\to \gamma\gamma$
decay.
This shows that the recent claim of decoupling by R. Gastmans, S.L. Wu, and T.T. Wu 
is incorrect, and the old calculations for the two photon decay well known in the literature are valid.
\end{abstract}
\vspace{1cm}

 \newpage
\noindent
 The two-photon mode of the Higgs-particle decay is important in experimental
searches. Therefore theoretical calculations of the $H\to \gamma\gamma$ rate  
received much attention. 
The $H$ transition into two photons goes via loops of charged particles: 
leptons, quarks and $W$ bosons. In the Standard Model (SM) the Higgs coupling to other fields
is proportional to the masses of the latter; the most massive particle have the strongest coupling.
All these loops had been calculated long ago \cite{EGN,IK,SVVZ,R}, by different methods, with totally consistent results. 

Surprisingly,  in two recent publications \cite{GWW1,GWW2} 
the issue of $H\to \gamma\gamma$ was raised anew, as if the passage of time negates the knowledge of the past.
Raymond Gastmans, Sau Lan Wu
and Tai Tsun Wu revisited the calculation of the $W$-boson loop in the $H\to \gamma\gamma$
decay claiming that the old results   \cite{EGN,IK,SVVZ,R} was erroneous. Using the unitary gauge they obtained a different  $H\to \gamma\gamma$ decay rate not coinciding with that of 
\cite{EGN,IK,SVVZ,R}. Technically, 
Gastmans et al. identify  dimensional regularization exploited in some previous
calculations as a source of the alleged mishap.

The main argument of Gastmans et al. in favor of the statement of \cite{GWW1,GWW2}  is the requirement of 
decoupling: their amplitude vanishes in the limit $m_W/m_H\,\to 0$ while that of \cite{EGN,IK,SVVZ,R} does not vanish 
in this limit. 

Superficially this argument might seem reasonable. Indeed, the above-mentioned decoupling works for the fermion loop
in the limit $m_f/m_H\,\to 0$ because the Higgs coupling to fermions is proportional to $m_f$.
Likewise,
in the $W$-boson case the Higgs coupling to $W^+W^-$  is proportional to $m_W^2$; thus
why not expect
vanishing of the $W$-loop contribution at $m_W=0$\,?

Actually this vanishing {\em does not occur}. In this note we will explain the {\em absence of decoupling} for the $W$-boson loop in the $m_W\to 0$ limit owing to some general features of the
non-Abelian vector fields. Our argument will connect a residual nonvanishing constant
in the $H\to \gamma\gamma$ amplitude at $m_W=0$ with a Goldstone-particle loop well known in the literature (see e.g. \cite{LS}).

There is a crucial difference between, say,  spin-1/2 and spin-1 particles
with regard to the massless limit. Namely, the number of polarization states stays the same for
spin-1/2 massive and massless particles,  while for the massive spin-1 particle we have three polarization states in contradistinction with the massless spin-1 field, with two polarization states.
In the massive case, in addition to two spatially transverse polarizations (intrinsic to the massless vector field) we have also the longitudinal polarization. Moreover, the amplitude of this polarization grows in the limit $m_W\to~0$. Indeed, the longitudinal polarization of the $W$ boson with 4-momentum $k^\mu=(E,0,0,k)$ moving along $z$-axis has the form
\begin{equation}
\label{eL}
\epsilon_L^\mu=\frac{1}{m_W}\big(k,0,0,E\big)=\frac{k^\mu}{m_W}+\frac{m_W}{E+k}
\big(-1,0,0,1\big)\,.
\end{equation}
In the case of an Abelian vector field the singular in $m_W$ term $\sim k^\mu/m_W$ does not contribute due to convolution with the conserved current; as a result 
the longitudinal quanta decouple in electrodynamics of a massive photon in the $m_\gamma\to 0$
limit.\footnote{\,Note, however, that even for Abelian vector (massive) fields the 
longitudinal polarizations do not decouple from gravity.}

For non-Abelian vector fields the longitudinal quanta do {\em not} decouple if two or more such  quanta are involved in the process under consideration. In terms of loops it means that starting from  one loop longitudinal $W$ bosons produce a nondecoupling contribution. This was first demonstrated as long ago as in 1971 in Ref. \cite{VKh}. 

In application to the $W$ loop in the $H\to \gamma\gamma$ decay we will show now that 
the longitudinal quanta do not decouple from the Higgs and, as a result,  the $W$- boson loop
does not vanish in the massless limit.

\vspace{2mm}

\noindent
In the Standard Model the Higgs coupling to $W$ bosons has the form
\begin{equation}
\label{HWW}
{\cal L}_{HWW}=2\,\frac{H}{v}\, m_W^2 W^\mu W^*_\mu\,,\qquad \frac{1}{v}=\big(G\sqrt{2}\big)^{1/2}.
\end{equation}
For our purposes the unitary gauge is sufficient for the
description of the $W$ vector field, the same gauge as was exploited in 
Ref. \cite{GWW1,GWW2}. There is no need to invoke  the $\xi$-gauge, ghost fields, {\em etc}.
Only the physical degrees of freedom are  of relevance. 

It is clear that the $HWW$ coupling following from the Lagrangian \eqref{HWW} vanishes
at $m_W=0$ for the transverse polarizations. The transverse polarizations can be safely neglected in this limit. At the same time 
the longitudinal polarization containing 
$1/m_W$ (as  seen from Eq. \eqref{eL}) remains coupled to $H$ 
(with a nonvanishing coupling) in the limit $m_W\to 0$. Technically,
we can substitute $W_\mu$ as follows
\begin{equation}
W_\mu \to W_\mu^L=\frac{1}{m_W}\,\partial_\mu \phi\,,\qquad m_W\to 0\,,
\end{equation}
where $\phi$ is a charged scalar field. Then,
\begin{equation}
\label{4}
{\cal L}_{HWW}\to 2\,\frac{H}{v}\, \partial^\mu \phi \partial_\mu \phi^*=
\frac{H}{v}\, \partial^2 \big( \phi \, \phi^*\big)
\end{equation}
where we neglected the terms
$\sim \partial^2 \phi$
which  is certainly perfectly legitimate on  mass shell. Equation (\ref{4}) means that
the Higgs field interacts with the Goldstone fields 
$\big( \phi \, \phi^*\big)$
through the trace of the energy-momentum tensor
of these Goldstone fields.

Furthermore, omitting total derivatives we can write
\begin{equation}
\label{HLL}
{\cal L}_{HWW}\to \frac{\partial^2 H}{v}\,  \big( \phi \, \phi^*\big)=
-\frac{m_H^2}{v}\, H\phi \, \phi^*\,.
\end{equation}

Summarizing, we have just demonstrated  that at $m_W=0$ the Higgs coupling to the longitudinally 
polarized $W$ bosons is equivalent to the coupling to the massless scalar field. This is nothing 
else but an abbreviated proof of the equivalence theorem at the tree level. A detailed derivation
of the equivalence one can find in Refs. \cite{VKh,CLT,CG}, see also \cite{KMY} for application
to the $H\to \gamma\gamma$ decay.

In the SM one can also readily establish directly from the Lagrangian that the coupling (\ref{HLL}) is the source of the $H \to \gamma \gamma$ amplitude in the limit $m_W \to 0$. Indeed, at a fixed $v$ and $m_H$, the latter limit corresponds to a vanishing $SU(2)$ coupling: $g \to 0$, so that the transversal $W$ bosons fully decouple from the Higgs boson. The only remaining relevant dynamics is that of the scalar sector,\footnote{\,This behavior, first established in Ref.\cite{VKh}, is also known as the equivalence theorem~\cite{CLT,CG}.} which in this limit contains the massive Higgs boson and three Goldstone bosons, two of which, $\phi^+$ and $\phi^-$, are charged and mediate the $H \to \gamma \gamma$ decay, due to the vertex described by Eq.\,(\ref{HLL}).

The next step, calculation
of matrix element $\langle \gamma \gamma|\phi \phi^*|0\rangle$ for transition of scalars to two photons, contains no ambiguity. For the massless scalars loop 
 we get \cite{SVVZ,LS,KMY} 
\begin{equation}
\langle \gamma \gamma|\phi \phi^*|0\rangle= -\frac{2}{(k_1+k_2)^2}\cdot \frac{\alpha}{4\pi}\,
(k_1^\mu e_1^\nu-k_1^\nu e_1^\mu)(k_{2\mu} e_{2\nu}-k_{2\nu} e_{2\mu})\,,
\end{equation}
where $k_{1,2}$ and $e_{1,2}$ are the  4-momenta and polarization vectors of the photons, respectively.
In the Higgs decay, with the coupling (\ref{4}),  $(k_1+k_2)^2=m_H^2$ and, therefore, 
\begin{equation}
\langle \gamma \gamma|{\cal L}_{HWW} |H\rangle= \frac{2}{v}\cdot \frac{\alpha}{4\pi}\,
(k_1^\mu e_1^\nu-k_1^\nu e_1^\mu)(k_{2\mu} e_{2\nu}-k_{2\nu} e_{2\mu})\,.
\end{equation}
At $m_W=0$  this coincides with results of Refs. \cite{EGN,IK,SVVZ}.
Thus, we demonstrate the origin of nondecoupling for the $W$-boson loop
at $m_W/m_H \to 0$ in the most transparent way. 

\vspace{3mm}

\noindent
 We would like add a few comments on the considerations in Refs. 
\cite{GWW1,GWW2}.

\vspace{2mm}

\noindent
{\bf (a)}
Instead of the two-photon decay let us consider first the decay $H\to W^+W^-$ in the limit $m_W\to 0$.
This is a tree-level process and its unambiguous calculation is sufficient to verify that (i) the amplitude does not vanish at $m_W=0$ and (ii) only longitudinal polarizations contribute in this limit,  in full correspondence with Eq.~\eqref{HLL}. It is seen where the argumentation of the authors of  
\cite{GWW1,GWW2} based on the requirement of decoupling  is flawed. It is an important point 
since Gastmans et al. understand that their diagrammatic expressions are not
well-defined {\em per se} in the unitary gauge. 

Moreover, one can start with the diagrammatic expression for the $W$ loop in the unitary gauge 
and apply the Ward identities for the singular in $m_W$ part coming from the vector propagators to verify the equivalence to the scalar loop at the small mass limit. This route was not explored in 
Refs. \cite{GWW1,GWW2}.
\vspace{2mm}

\noindent
{\bf (b)}
The authors of \cite{GWW1,GWW2} suggest that an inaccurate use of dimensional regularization causes
the difference between their results and  results of \cite{EGN,IK,SVVZ,R}. While this regularization had been indeed used in Ref. \cite{EGN}
we did not invoke it at all in our old calculation \cite{SVVZ}. 
And yet, the results of \cite{EGN} and  \cite{SVVZ} are in accord with each other.
Moreover, the absence of decoupling is visible
at the tree level as was noted in the previous comment. Obviously,  one cannot then blame dimensional regularization of the one-loop integrals. 

\vspace{2mm}

\noindent
{\bf (c)}
The low-energy theorem we derived in \cite{SVVZ} relates the $H\gamma\gamma$ amplitude
in the opposite limit $m_W/m_H \to \infty$ to the $\beta$ function of the corresponding particles.
For the massive $W$ bosons this $\beta$ function was first found in 1965 by Vanyashin and Terentev 
\cite{VT};  actually,  it was the first signal of the soon-to-be discovered asymptotic freedom. Their calculation is 
simple to verify by modern methods.\footnote{\,The $\beta$-function coefficient is $b=7$ for 
$W$ bosons, see \cite{SVVZ} for details.
It worth noting that in \cite{VT} one can read off this coefficient from the large field strength limit.
In the charge renormalization Vanyashin and Terentev had $b=20/3=7-1/3$. The extra $-1/3$ 
should be taken away, it comes from an auxiliary heavy scalar with the mass $m_W/\sqrt{\xi}\,,~\xi \to 0$. The parameter $\xi$ in \cite{VT} is just an inverse of $\xi$ in the $R_{\xi}$ gauge where the $\xi$ dependence is canceled by ghosts. }
  This gives an independent
check of calculations \cite{EGN,IK,SVVZ} and contradicts to \cite{GWW1,GWW2}.

\vspace{2mm}

\noindent
In summary. We reconfirm the results of the previous calculations\cite{EGN,IK,SVVZ,R} of the $H \to \gamma \gamma$ decay amplitude, including the nondecoupling of the Higgs boson from the two photon channel in the limit $m_W/m_H \to 0$ as found (and emphasized) in our old calculation~\cite{SVVZ}. We have explicitly demonstrated here that this behavior is due to the contribution of the longitudinal $W$ bosons in the intermediate state and can be traced to the Goldstone modes of the scalar field within the context of the well known equivalence theorem. We thus assert that the recent claim~\cite{GWW1,GWW2} of an error in previous calculations is incorrect.

\vspace{3mm}

\noindent
 We would like to acknowledge helpful discussions with John Collins, Bernd Jantzen, Kirill
Melnikov, and Misha Vysotsky. The work of M.S., A.V., and M.B.V. is supported in part by the DOE grant DE-FG02-94ER40823.

\vspace{3mm}

\noindent
{\bf Note added:} After the ArXiv posting  of our text two more papers \cite{HTW,MZW}
on the same subject were posted. The detailed analysis given in these papers identifies
a culprit in the calculations \cite{GWW1,GWW2}: they do not maintain the electromagnetic 
gauge invariance. Once this invariance is supported, either by regulators of Pauli-Villars type 
\cite{HTW}, or by dimensional regularization \cite{MZW}, the correct result is unambiguous.

To pinpoint the issue let us consider the integral
\begin{equation}
\label{8}
I_{\mu\nu}=\int \!\!{\rm d}^{4} l \,\frac{4l_{\mu}l_{\nu}-g_{\mu\nu}l^{2}}{\big(l^{2}-B+i\epsilon\big)^{3}}\,.
\end{equation}
which appears in the calculations \cite{GWW1,GWW2}. 
The authors put this integral to zero. It is true indeed under spherically symmetric integration over $l$ (after Euclidian continuation) which they have used. On the other hand, the integral (\ref{8}) can be rewritten
in the form
\begin{equation}
\label{9}
I_{\mu\nu}=\frac{1}{2}\,\int \!\!{\rm d}^{4} l \,\frac{\partial^{2}}{\partial l^{\mu}\partial l^{\nu}}\,\frac{1}{l^{2}-B+i\epsilon} -g_{\mu\nu}\int \!\!{\rm d}^{4} l \,\frac{B}{\big(l^{2}-B+i\epsilon\big)^{3}}\,.
\end{equation}
The second integral is well defined and gives just a number $\int \!\!{\rm d}^{4} l \,{B}\,\big(l^{2}-B+i\epsilon\big)^{-3}=-i\pi^{2}/2$\,. The first integral of total derivatives which cancels this number 
in dimension 4 is the one which breaks gauge invariance. Indeed, it can be viewed as a second order
of expansion in the constant gauge potential $A_{\mu}$ of the expression
\begin{equation}
\label{Aconst}
\int {\rm d}^{4} l \,\frac{1}{(l+eA)^{2}-B+i\epsilon}\,.
\end{equation}

Thus, to preserve gauge invariance we should put the integral of total derivative to zero.
This is automatic in both, dimensional regularization and Pauli-Villars one, but we can also 
use as a constraint which allows to maintain the current conservation. 

Once it is realized that the integral (\ref{8}) is nonvanishing, $I_{\mu\nu}= g_{\mu\nu} i\pi^{2}/2$\,,
it is simple to check that substituting this into GWW calculations leads to reproducing of the standard result. It appears first in terms $1/m_{W}^{2}$, Eq.\,(3.40) of Ref.\cite{GWW2}, where instead of vanishing 
one arrives at a nonvanishing at large $m_{H}$ term. The integral (\ref{8}) also appears in 
Eq.\,(3.50) of Ref.\cite{GWW2} for nonsingular in $m_{W}$ terms. The above substitution leads to the transversal result (3.52) without any  kind of subtraction, no Dyson prescription is needed. 

The generic issue of finite but undetermined loops was discussed earlier by 
Roman Jackiw \cite{RJ}. We thank him for pointing this out to us.


\newpage
\begin{thebibliography}{99}

\small

\bibitem{EGN}
J.~R. Ellis, M.~K. Gaillard, and D.~V. Nanopoulos, Nucl.\ Phys.\ B {\bf 106}, 292 (1976).
\bibitem{IK}
B.L. Ioffe and V.A. Khoze, Sov.\ J.\ Part.\ Nucl.\ {\bf 9}, 50 (1978) \\[0mm]
[Fiz.\ Elem.\ Chast.\ Atom.\ Yadra {\bf 9}, 118 (1978)]. 
\bibitem{SVVZ}
M.A. Shifman, A.I. Vainshtein, M.B. Voloshin, and V.I. Zakharov, \\[0mm] Sov.\ J.\ Nucl.\ Phys.\ {\bf 30},711 (1979) [Yad.\ Fiz.\ {\bf 30}, 1368 (1979)].
\bibitem{R}
T.~G. Rizzo, Phys.\ Rev.\ D {\bf 22}, 178 (1980).
\bibitem{GWW1}
R.~Gastmans, S.~L.~Wu, and T.~T.~Wu, 
 ``Higgs Decay $H \to \gamma\gamma$ through a $W$ Loop:\\[0mm] Difficulty with Dimensional Regularization,''
arXiv:1108.5322 [hep-ph].
\bibitem{GWW2}
R.~Gastmans, S.~L.~Wu, and T.~T.~Wu,
 ``Higgs Decay into Two Photons, Revisited,''
arXiv:1108.5872 [hep-ph].

\bibitem{LS}
H.~Leutwyler and M.~A.~Shifman,
  Phys.\ Lett.\  {\bf B221}, 384 (1989).

\bibitem{VKh}
A.I. Vainshtein and I.B. Khriplovich, 
Sov. J. Nucl. Phys., {\bf 13}, 111 (1971)\\[0mm] [Yad.\ Fiz.\  {\bf 13}, 198 (1971)].

\bibitem{CLT}
 J.~M.~Cornwall, D.~N.~Levin, and G.~Tiktopoulos,
  Phys.\ Rev.\  {\bf D10}, 1145 (1974); (E) {\bf 11}, 972 (1975); 
  C.E. Vayonakis, Lett. Nuovo Cim. {\bf 17}, 383 (1976).
  
\bibitem{CG}
M.~S.~Chanowitz and M.~K.~Gaillard,
  Nucl.\ Phys.\  {\bf B261}, 379 (1985). 
 
\bibitem{KMY}
J.~G.~Korner, K.~Melnikov, and O.~I.~Yakovlev,
  Phys.\ Rev.\  {\bf D53}, 3737-3745 (1996) [hep-ph/9508334].
    
\bibitem{VT}  
V.S. Vanyashin and M.V. Terentev,
Sov. Phys. JETP, {\bf 21}, 375 (1965)\\[0mm] [J. Exptl. Theoret. Phys., {\bf 48}, 565 (1965)].

\bibitem{HTW}
D.~Huang, Y.~Tang, and Y.-L.~Wu,
``Note on Higgs Decay into Two Photons $H\to \gamma\gamma$,''
arXiv:1109.4846 [hep-ph].

\bibitem{MZW}
W.~J.~Marciano, C.~Zhang, and S.~Willenbrock,
``Higgs Decay to Two Photons,''
  arXiv:1109.5304 [hep-ph].
  
\bibitem{RJ}  
R.~Jackiw,
  Int.\ J.\ Mod.\ Phys.\  {\bf B14}, 2011 (2000)
  [hep-th/9903044].

\end{thebibliography}
\end{document}